# Coherent control of photomagnetic back-switching by double-pump laser pulses

T. Zalewski[*], L. Nowak, and A. Stupakiewicz

*Faculty of Physics, University of Bialystok, 1L Ciolkowskiego, 15-245 Bialystok, Poland*
*Corresponding author: tmkzlwsk@uwb.edu.pl*

**The control of nonthermal, all-optical magnetization switching under the regime with an independent state of laser polarization opens up new opportunities for ultrafast magnetic recording. Here, we investigate the photo-magnetic back-switching capabilities of the write and erase magnetic domain pattern using double-pump pulse excitations in an iron garnet film with pure cubic magnetocrystalline symmetry. It is essential to note that forward and backward magnetization switching is achievable in two distinctive scenarios: using identical linearly polarized laser pulses and with pulses having orthogonal polarization planes. By observing the switch of magnetization at domains independent of the initial state, one can nonthermally toggle the magnetization, equivalent to XOR logical operation, at frequencies reaching up to 50 GHz.**

In the past decade, several mechanisms, finally leading to the ultrafast all-optical switching (AOS) of magnetization in both metals and dielectric materials have been shown [1]. Qualitatively, the magnetization switching solely using the laser pulse can be either non-deterministic – every incident pulse can switch the magnetization resulting in toggle switching or deterministic – the polarized laser pulse introduces some kind of break in the symmetry, which determines the switching between two states of magnetization. In the case of the thermal-based ultrafast AOS in metallic materials [2,3], both regimes of magnetization switching were already demonstrated [4–6]. For instance, for the deterministic switching the degeneracy between magnetic states can be broken by the helicity of circularly polarized light. In such all-optical helicity-dependent switching (AO-HDS) [2,6–8] the final state of the magnetization is controlled and determined by the left or right circular polarization of the optical pulse. On the other hand, in specific materials such as Gd-based alloys, multi-sublattice alloys such as Gd/Co [9], or synthetic multilayers with nanostructures of Gd/Co [10], Tb/Co [11,12] the magnetization can be switched only by a single femtosecond pulse regardless of its polarization state. This all-optical helicity independent switching (AO-HIS) [3,10,11] gives a toggle effect making the final magnetization state dependent on the parity of incident pulses.

In the case of dielectric materials, ultrafast magnetization control can be achieved through thermal effects [13]. However, when based on the photo-magnetic effect, it does not entail a heat load, making the switching a non-thermal process. It has been experimentally shown that in yttrium iron garnet doped with cobalt (YIG:Co) [14], it is possible to permanently switch the magnetization state within only 20 ps by using a single ultrashort linearly polarized laser pulse. Light pulse polarization along [100] or [010] axis resonantly excites specific $d-d$ transition in Co-ions modifying the magnetic anisotropy. This photo-induced anisotropy can lift the degeneracy between the domain states, thus permanently switching the magnetization. This has been demonstrated in YIG:Co films with a 4° miscut angle, showing that using the femtosecond laser pulse can write, erase and re-write magnetic domains only after changing the orientation of linear polarization at frequencies up to 20 GHz [15].

The determination of possible frequencies and repetition times for controlling the magnetization state can be directly obtained by employing double-pump pulses, separated by a defined time delay. This method is recently commonly applied in the case of metallic ferrimagnets, such as GdFeCo [16,17] or TbFeCo [18], where the recovery time after thermally induced changes is crucial. Reducing this recovery time accelerates writing and erasing cycles [19,20]. This approach is also relevant for half-metallic Heusler ferrimagnet $Mn_2Ru_{0.9}Ga$ [21], as well as in the case of AO-HDS in Pt/Co/Pt [22]. Furthermore, the subsequent optical pulse provides an additional degree of freedom in controlling magnetization dynamics through different mechanisms, such as the coherent control of spin waves in ferrimagnetic garnets [23] or the magneto-acoustic mechanism driven by quasi-static strain [24].

Recently, the nonthermal toggle switching regime observed in YIG:Co films with pure magnetocrystalline cubic symmetry [25] was shown. Here for distinction from the thermal-based toggle switching we are referring to this regime as the back-switching of magnetization. In this regime, the magnetization state can be controlled solely by the sequence of femtosecond laser pulses without changing their polarization direction. The control of the back-switching regime with the precessional mechanism of magnetization switching raises several new questions. Firstly, if one pulse can switch magnetization and a subsequent pulse can reverse it, what will happen if both of these pulses appear together? On the other hand, from an application perspective, where the change in the direction of magnetization from 'down' to 'up' corresponds to the bit state '1', and the reverse corresponds to '0', the lack of necessity to control the laser polarization state for writing and erasing magnetic 'bits' can significantly simplify the device. This eliminates the need for changing the laser polarization at high frequencies to match the recording speed.

We used the YIG:Co thin films with the composition $Y_2CaFe_{3.9}Co_{0.1}GeO_{12}$ and a thickness of 8 μm. The garnet film was grown using the liquid phase epitaxy method on the gadolinium gallium garnet ($Gd_3Ga_5O_{12}$) on (001)-plane substrate with miscut <0.1°. The dominant negative cubic magnetic anisotropy $K_c$ = -5.5×10$^3$ erg/cm$^3$ with small uniaxial contribution with constant $K_u$= 0.6×10$^3$ erg/cm$^3$ results in energetically equivalent four easy magnetization axes along the cube diagonals of <111>-type directions. The pure cubic magnetocrystalline symmetry domain structure observed in magneto-optical microscopy is balanced, with magnetic domains covering equivalent volumes and forming a uniformly distributed labyrinth pattern.

To observe the magnetic domain pattern and magnetization switching we utilized the magneto-optical microscope in Faraday geometry. The magnetization dynamics of photo-magnetic switching were examined using the time-resolved pump and probe technique [26] utilizing a double-pump excitation scheme. [15,17] As a source of laser pulses the used setup utilized an ultrafast Ti:sapphire laser system, that provides a train of pulses with a duration of 35 fs, wavelength of 800 nm, and base repetition rate of 1 kHz. Its beam was split into three separate branches. One of them, maintaining the central 800 nm wavelength was used as a probe beam. Two pumps, referred to as pump A and pump B, were directed to separate optical parametric amplifiers, enabling wavelength conversion and selecting the 1300 nm wavelength. Next, the beams were directed toward separate mechanical delay lines, which allowed for adjusting the mutual delay between the first pump A and the second pump B set as $\Delta t_{AB}$ time. The probe beam illuminated the sample at normal incidence and two pump beams were directed at an angle of about +/-10° from the normal incidence. All measurements were performed without an external magnetic field and at room temperature.

Since the back-switching can be observed in the statics regime, in which the exposure time enabling obtaining magneto-optical images can be arbitrarily long and last several hundred milliseconds, in a relatively simple experiment using double pump excitation, we were able to analyze pump pulse pair interference. First, we precisely adjusted the optical parameters of pumps A and B to match each other. The pump fluence (50 mJ/cm$^2$), polarization plane along to [100] crystallographic orientation in YIG:Co film, and the localization on the sample were unified to provide the exact switching area using either A or B a single laser pulse, whose size is marked in the first image in Fig. 1a. The presented differential structure represents switching between [11-1] and [1-11] magnetization directions. The initial state depends on the differentiating image selection. The orientation and the presence of all four different domain phases with magnetization along <111>-type directions were verified by tilting the sample from the plane normal by an approximately 15° angle, allowing for the simultaneous visualization of out-of-plane and in-plane magnetization components in the magneto-optical Faraday geometry.

Next, the sample was excited by both pulses A and B being in spatio-temporal overlap. By delaying the pulse B and increasing $\Delta t_{AB}$, a full set of double pump images was captured.

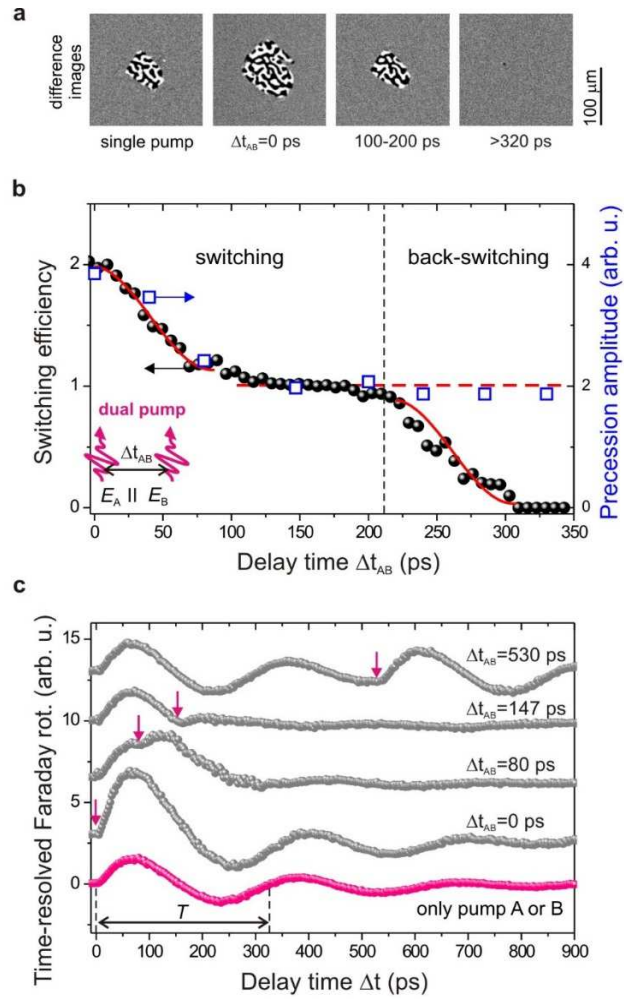

FIG. 1. (a) Magneto-optical difference images presenting a double pulse writing sequence with two identical pump pulses A and B with polarization $E_A$ and $E_B$ along the [100] axes and delayed by $\Delta t_{AB}$. The first image shows the switching after excitation with either a single A or B pump pulse. Subsequent images present the obtained switching pattern during double pump pulse excitation with the marked delay $\Delta t_{AB}$. (b) The normalized switched area as a function of delay time $\Delta t_{AB}$ (circled black points). The solid red line directing to the saturation regions was fitted using $\cos(2\pi\Delta t_{AB}/t_{sw})$, where $t_{sw} = 178.6 \pm 5.2$ ps is the half period

of magnetization precession (*T*) [14]. The square blue points represent the amplitude determined from the double pump traces as the amplitude of magnetization precession. The dashed red line is a guide for the eye. (c) The time-resolved Faraday rotation signal from the only single pump pulse (red points) and the double pump and probe pulses for different times $\Delta t_{AB}$. Both pump pulses were polarized along the same polarization plane along [100] direction in YIG:Co. The curves are offset vertically. The arrows mark the temporal position of pump pulse B, which is delayed by $\Delta t_{AB}$.

To quantitatively analyze the impact of the second pump pulse on switching, we expressed the efficiency of switching as a normalized switching area to area for single pump pulse switching. Because determining the normalized switching area in the case of a labyrinth-like structure may not be obvious, we used the same repeatable procedure for every analyzed image. Firstly, the analyzed image was prepared in differential form. Next, to remove any remaining background non-uniformities, the image was once again differentiated with a copy blurred in a Gaussian filter and subjected to bandpass filtration. Such an image was separated into two parts with a low threshold and a high threshold, selecting a particular part (with white or black contrast) of the domain structure. Next, the images were added together once again and binarized. Finally, after a set of binary operations, including filling holes and closing the structure, the switching size was determined and could also be translated into switching efficiency. During the double-pump excitation, the magnetization switching occurs up to $\Delta t_{AB}$ = 320 ps. The dependence of the switching area for $\Delta t_{AB}$ time is depicted in Fig.1b. This obtained time-dependent trace is characterized by three zones. To understand this shape, measurements involving double-pump time-resolved measurements with pulses fluence 10 mJ/cm$^3$ are required.

In order to conduct these measurements, we utilized a double pump and probe setup capable of time-resolved tracking of magnetization precession for pump laser fluence (<10 mJ/cm$^2$) below the switching threshold. In this configuration, the probe beam, operating at a 1 kHz repetition rate and focused on the sample with a 50 μm diameter polarized orthogonally to the pump pulse polarization was employed. In order to detect temporal changes in the Faraday rotation signal, which is proportional to the magnetization component ($M_z$) along [001] crystallographic direction in the sample, a scheme with an auto-balanced photodiode was used. After traversing the sample, the probe beam was directed to a half-wave plate and a Wollaston prism, which separated it into two linearly polarized beams with orthogonal polarization. These beams were directed to two separate branches of an auto-balanced photodiode, which monitored the change in their intensities corresponding to Faraday rotation. In this case, for every particular delay between pulses A and B, the probe beam was swept with the delay $\Delta t$, revealing both pump beams' impact on magnetization precession. Both pump pulses were once again equalized and matched to individually induce exactly the same precession signal.

The obtained time-resolved double pump traces are presented in Fig. 1c, with arrows marking the position of pump pulse B. We note that for $\Delta t_{AB}$ = 0 ps, the mutual overlapping of the laser pulses causes constructive interference, enhancing the effective amplitude of precession. Therefore, the overlapping of two identical laser pulses increased the amplitude of photo-magnetic precession and switching area (see Fig.1b) by a factor of two. The precession amplitude was determined using an exponentially damped sine function to fit the data. The superposition of the pump laser pulses effectively increases the amplitude, reaching its maximal value at $\Delta t_{AB}$ = 0 and decreasing to the level obtained by a single pulse in approximately $\Delta t_{AB}$ = 75 ps. This correlates with the region of decreased switching efficiency. Further, the two signals of magnetization precession start interfering with each other in counter-phases, resulting in the freezing of the magnetization precession after half the period $T$ for $\Delta t_{AB}$ ~ 147 ps (see Fig. 1c). In this range, the effective amplitude is defined solely by pulse A, and the switching efficiency does not change up to about $\Delta t_{AB}$ = 210 ps (see Fig. 1b) demonstrating the logical operation AND. Finally, for $\Delta t_{AB}$ > 210 ps, the laser pulses are separated far enough not to be able to constructively interfere with subsequent sine-like precession sufficiently to overcome the switching threshold, as a single effective pulse. Therefore, in the switching regime for pump laser pulses with a fluence of 50 mJ/cm$^3$, while the first pulse A provides switching, the second B, starting from a position with already almost stopped magnetization motion, provides a separate stimulus resulting in back-switching. Such a regime is equivalent to XOR logical operation. The complete back-switching occurs at $\Delta t_{AB}$ > 320 ps, which can be correlated with the magnetization precession period of frequency. In contrast, as observed in the same composition of YIG:Co film with miscut, the switching state after the second pulse remained stable above 200 ps. Such observation further confirms the amplitude dependence (dashed line in Fig. 1b), suggesting that after this time, a sequence of pulses remains constant, and the subsequent state will be the same as the initial one. It is worth noting that pulse fluence significantly influences both the switching efficiency and dynamic measurements. For instance, as demonstrated in [15] mutual interference of two pump fluences below the switching threshold can surpass the threshold and provide switching. Moreover, very high fluences could potentially allow for interference between further oscillations maxima, resulting in more "steps" in the switching area dependence in Fig.1c.

To demonstrate the repeatability of the photo-magnetic switching within a single pump pulse in a picosecond time scale, we captured a series of single-shot time-resolved images using the setup described in detail in Ref. [26]. The wavelengths and polarizations of laser pulses were identical to those specified for pump and probe dynamics. However, the probe beam was defocused to a diameter larger than 500 μm to illuminate a sufficiently large area for imaging. The time-resolved magneto-optical image, generated by the rotation of the plane of polarization induced by the pump pulse, was captured on a synchronized CCD camera.

The exemplary stack of time-resolved differential images is presented in Fig. 2a, with the temporal evolution of the magnetic contrast, proportional to the normalized out-of-plane magnetization component $\Delta M_z$, shown in Fig. 2b. In this experiment, we did not apply an external resetting magnetic field, as was done previously in the YIG:Co sample with miscut. Unexpectedly, we achieved magnetization reversal from [11-1] to [1-11] directions and vice versa without resetting. This confirms the previously observed toggle-switching regime [25], in which the end state of the switching becomes the initial state for the subsequent pulse. It is evident that the contrast develops rapidly for 20-90 ps. However, as in the case of samples with a miscut angle, the switching trajectory within different domains is characterized by asymmetry in two types of trajectories: towards M+ and M- magnetization states along [1-11] and [11-1] directions, respectively (see Fig. 2b) [27].

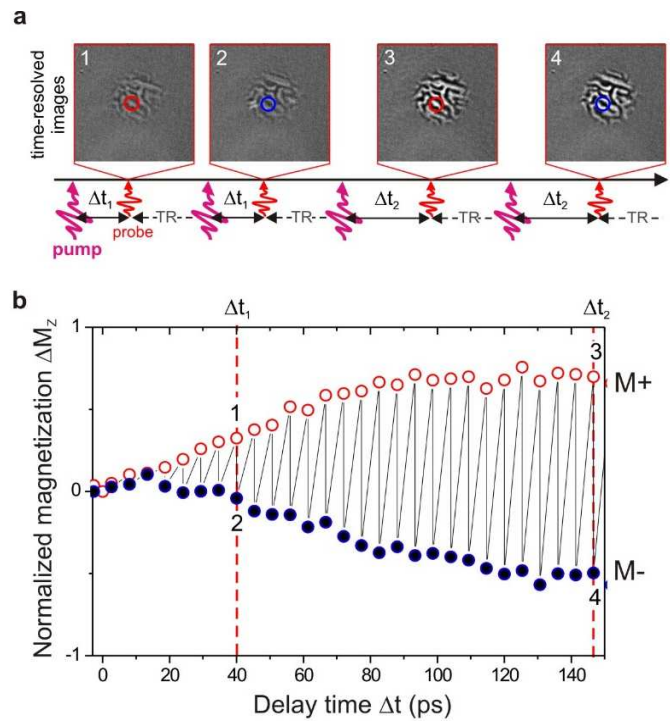

FIG. 2. (a) The differential single-shot time-resolved images of the magnetic pattern obtained at $\Delta t_1$ = 40 ps and $\Delta t_2$ = 147 ps. Each subsequent image starts as an initial state for the next pump pulse switching. The observed change in the image intensity corresponds to the change in the normalized magnetization component. Between every captured time-resolved image the $TR$ < 50 ms time required for the data acquisition, therefore the initial state is always fully switched. (b) The time-resolved change of the Faraday rotation proportional to the change in the out-of-plane magnetization component $\Delta M_z$. The traces were retrieved from the image stack from the region marked by colored squares separately for both black with M- along [11-1] and white M+ along [1-11] directions in domains.

In order to demonstrate a resetting scenario we utilize two pump pulses polarized orthogonally to each other. Using the same approach and settings, we performed analogous measurements, only changing the polarization plane of the pump pulse B. The obtained measurements are provided in the same manner, with the exemplar stack of differential images of domain pattern presented in Fig. 3a, normalized switching area dependence utilizing double pump excitation presented in Fig. 3b, and double pump and probe time-resolved magnetization dynamics in Fig 3c.

With pump A polarized along the [100] axis and pump B polarized along the [010] axis, the generated torque has opposite signs. Therefore, for $\Delta t_{AB}$ = 0 ps, precessional signals being in counter phases interfere destructively, resulting in no magnetization precession and therefore do not provide any switching. In contrast to previous results on the YIG:Co miscut sample, this confirms the existence of pure cubic symmetry and energetically equivalence of direction of magnetization during forward and backward switching. However, by increasing $\Delta t_{AB}$, a characteristic region between $\Delta t_{AB}$ = 30 - 80 ps in which switching is obtainable occurs. Once again, to understand this behavior, insight into dynamics is required. As it may be noticed, when precession traces are delayed, the interference changes its character and starts to provide precession similar to the one obtained by a single pump. Further increase of $\Delta t_{AB}$ indeed increases the amplitude of precession to the saturated level, however, after about a quarter of the precession period $T/4$, the second pump pulse B is able to overcome switching. It results once again in switching with pulse A and magnetization reversal with pulse B.

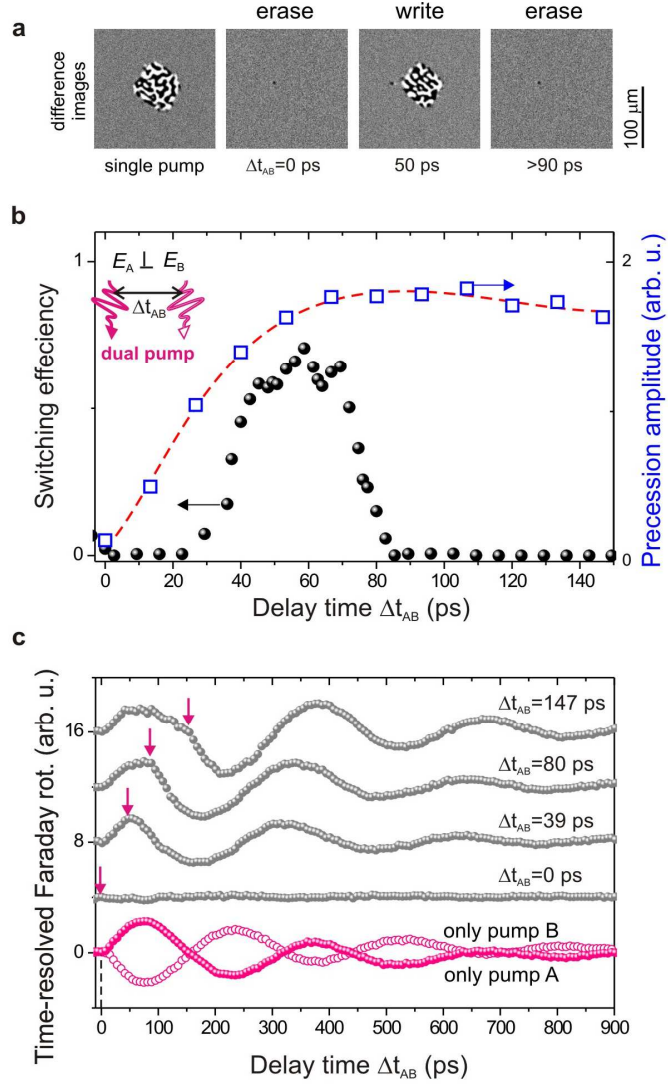

FIG. 3.(a) Magneto-optical difference images presenting a double pulse writing sequence with two pulses A with $E_A \parallel [100]$ and B with $E_B \parallel [010]$ axis. As previously, the first image shows the switching after excitation with either a single A or B pump pulse with the size used for normalization. The subsequent image represents the obtained switched pattern depending on the $\Delta t_{AB}$. (b) The normalized switched area as a function of delay time $\Delta t_{AB}$ (circled black points). The square blue points represent the amplitude determined from the double pump traces as the amplitude of magnetization precession. The dashed line represents the effective precession amplitude of both pump pulses. (c) The time-resolved Faraday rotation signal from the double pump and probe setup with pump A polarized along [100] and pump B along [010] planes. Notably, the two pump signals, being in counter-phase, interfere completely destructively $\Delta t_{AB} = 0$ ps.

These results underline that the photo-magnetic back-switching regime with the precessional motion of magnetization opens a plethora of possibilities for writing and erasing magnetic bits using only a sequence of the ultrashort laser pulses. In this regime, the lifetime of photoinduced anisotropy is limited to approximately 20 ps and is independent of the initial state of magnetization with a fixed laser polarization state. Therefore the magnetization state can be toggled for laser pulses with control of the delay time between pulses, achieving a switching frequency of up to 50 GHz. The toggle switching regime can also be successfully used to perform basic logical operations, such as XOR, AND, or inverting operations equivalent to the NOT operation. In fact, the utilization of only a set of connectives AND and NOT provides functional completeness, giving access to any other logical operation, as every switching function can be expressed by means of operations within it. Supporting and controlling magnetization switching, for example, both laser and electric pulses, could provide an additional degree of freedom in the modification of extrinsic contribution to magnetic anisotropy in the photo-magnetic garnets. This potential advancement could find utility in emerging fields such as opto-spintronics or opto-magnonics, utilizing the multi-states character of cubic garnet systems and offering opportunities for logical operations.

**Acknowledgments.** Work supported by Foundation for Polish Science (POIR.04.04.00-00-413C/17-00) and the European Union's Horizon 2020 Research and Innovation Programme under the Marie Skłodowska-Curie grant agreement No 861300 (COMRAD). We thank Prof. A. Kimel and Prof. A. Maziewski for the fruitful discussions. The data that support the findings of this study are available from the corresponding author upon reasonable request.